\documentclass[preprint,amsmath,amssymb,nofloats,aps]{revtex4}
\usepackage{graphicx}
\usepackage{multirow}
\usepackage{dcolumn}
\usepackage{bm}
\usepackage{amssymb}

\begin{document}

\title{NANOELECTRONICS}
\author{G. Allan$^{*}$, C. Delerue, C. Krzeminski}
\affiliation{Institut d'Electronique et de Micro\'electronique du Nord, D\'epartement Institut Sup\'erieur d'Electronique du Nord, 41 boulevard Vauban, 59046 Lille C\'edex, France}
\author{M. Lannoo}
\affiliation{Laboratoire Mat\'eriaux et Micro\'electronique de Provence, ISEM, Place G. Pompidou, 8300 Toulon, France}

\email{guy.allan@isen.fr}

\maketitle

\section{Introduction}
In this chapter we intend to discuss the major trends in the evolution of microelectronics and its eventual transition to nanoelectronics. As it is well known, there is a continuous exponential tendency of microelectronics towards miniaturization summarized in G. Moore's empirical law. There is consensus that the corresponding decrease in size must end in 10 to 15 years due to physical as well as economical limits. It is thus necessary to prepare new solutions if one wants to pursue this trend further. One approach is to start from the ultimate limit, i.e. the atomic level, and design new materials and components which will replace the present day MOS (metal-oxide-semiconductor) based technology. This is exactly the essence of nanotechnology, i.e. the ability to work at the molecular level, atom by atom or molecule by molecule, to create larger structures with fundamentally new molecular organization. This should lead to novel materials with improved physical, chemical and biological properties. These properties can be exploited in new devices. Such a goal would have been thought out of reach 15 years ago but the advent of new tools and new fabrication methods have boosted the field. We want to give here an overview of two different subfields of nanoelectronics. The first part is centered on inorganic materials and describes two aspects: i) the physical and economical limits of the tendency to miniaturization; ii) some attempts which have already been made to realize devices with nanometric size. The second part deals with molecular electronics, where the basic quantities are now molecules, which might offer new and quite interesting possibilities for the future of nanoelectronics.

\section{Devices built from inorganic materials}
These are mainly silicon based microelectronic devices which have invaded our life. Integrated circuits are now found everywhere not only in Personal Computers but also in a lot of equipment we use each minute as cars, telephones, etc. We always need more memory as well as faster and cheaper processors. 
The race for miniaturization began just after Kahng and Atalla \cite{ref1} demonstrated in 1960 the first metal-oxide semiconductor field effect transistor (MOSFET) (Fig. \ref{fig:fig1}). It turned out to be a success because a large number of transistors and their interconnections could be easily built on the surface of a single silicon chip. Ten years later, the first 1 kilobyte memory chip was on the market and this trend has been followed until now (a 64 megabit one contains more than one hundred millions electronic components). In 1965, Gordon Moore predicted what is known as the Moore's law: for each new generation of memory chip on the market, the number of components on a chip would quadruple every three years. Miniaturization not only decreases the average current cost per function (historically, $\sim~$ 25\% /year) but also improves the cost-performance ratio. In the same time, the market growth was close to 15\%/year. 
Most of the improvement trends are exponential and are resumed in the scaling theory \cite{ref2}. It shows that a MOSFET operates at a higher speed without any degradation of reliability if the device size is scaled by a factor $1/k$ and at the same time the operating voltage is scaled by the same factor. The speed of the circuits has increased up to one Gigahertz in today's personal computers. Continued improvements in lithography and processing have made possible the industry's ability to decrease the minimum feature sizes used to fabricate integrated circuits. For four decades, the semiconductor technology has distinguished itself by the rapid pace of improvement in its products and many predicted technological limitations have been overcome. In the same time it is difficult for any single company to support the progressively increasing R\&D investments necessary to evolve the technology. Many forms of cooperation have been established. as the «International Technology Roadmap for Semiconductors" \cite{ref3} which is the result of a large world consensus among leading semiconductor manufacturers. It looks at the 10-15 years in the future showing that most of the known technological capabilities will be approaching or have reached their limits. Continued gains in making VLSI (Very Large Scale Integration) seem to be first severely limited by technological considerations \cite{ref4}. 
The first one which is known as a «power crisis» is an heat dissipation problem. While the size of the components is reduced by a factor $k$, the scaling theory shows that the power density increases like $k^{0.7}$ \cite{ref2}. A today's single chip processor in production requires 100 W and this value will rise to 150 W during the next decade. Low-power design of VLSIs is also necessary because they are more and more used in mobile electronic systems which need long-lasting batteries. The reduction of the supply voltage ($\sim$ 0.37 V in 2014) will be accompanied by an increase of the current needed to operate VLSIs to huge values (500 A). This contributes to the second limitation: the «interconnect crisis».
A large operating current gives rise to voltage drop problems due to the resistance of the interconnections and to reliability degradation due to electro-migration of defects in the conducting wires. On the other hand, to be attractive, a scaling of the components must be accompanied by a scaling of the interconnect line thickness, width and separation. Then signal integrity is also becoming a major design issue. A high crosstalk noise is due to larger capacitive couplings between interconnects. A smaller geometry also increases the RC (resistance-capacitance) delay (it increases as k$^{1.7}$ \cite{ref2}). If this delay increases, the signal cannot propagate anywhere within the chip within a clock cycle.
According to T. Sakurai \cite{ref2}, a «complexity crisis» will appear. Design complexity is increasing superexponentially. The first obvious reason is the increased density and number of transistors. The complexity is also growing due to designs with a diversity of design styles, integrated passive components and the need to increase incorporate embedded software. The integrated circuit is built on several interconnected levels which interact. Verification complexity grows with the need to test and validate the designs. Finally the tests must be done at higher speed, higher levels of integration and greater design heterogeneity.
There are potential solutions to solve some of these problems during the 15 next years \cite{ref3}. In some other cases the solution is still unknown and this will have a price that people could not continue to afford. Moreover until now, we have considered more technological limitations but new fundamental quantum phenomena will also appear. The first one which has already been investigated is the tunnel effect through the gate insulator which is made of the silicon native oxide SiO$_{2}$ \cite{ref4}. One important factor of the MOSFET is the gate capacitance of the parallel plate capacitor made by the gate and the conducting channel in the semiconductor (Fig. \ref{fig:fig1}). This capacitor is filled by an insulator which is at present silicon dioxide. The charge of this capacitor controls the current between the drain and the source. When the gate (which is one of the capacitor plates) is scaled by $1/k$, the gate thickness scales as $1/k$ to maintain the same capacitance. It will be reduced to 0.7 nm in 2014. at the same time the gate voltage should be reduced to maintain the electric field across the oxide below an undesirable value. This is equivalent to 3 layers of oxygen and 2 of silicon in the oxide. When the insulator thickness is reduced, the overlap of the electronic wavefunctions on both barrier sides increases exponentially and this gives rise to a non-zero probability for the electrons to tunnel across the barrier. An other formidable challenge is to replace the silicon oxide by a dielectric material with a higher dielectric constant. A review of current work and literature in the area of alternate gate dielectrics is given in reference \cite{ref5}. 
Doping the semiconductors with impurities, donors (n type) or acceptors (p type), is an essential feature to get free carriers and thus get a substantial conductivity. With current impurity concentration one can show that below a $0.1 * 0.05$ mm gate, there are about 100 free carriers only. This means that a fluctuation of plus or minus one charged impurity in the channel gives a 1\% error, which represents another type of limitation. Some other new quantum effects appear due to the size reduction:

\begin{itemize}
\item{as in a molecule or an atom, the electron energy can only take discrete value as opposed to the classical value or to the existence of bands of allowed energies for bulk materials.}

\item{when the wave function associated with an electron takes several distinct channels, interference effects occur which give rise to conductance fluctuations with a root mean square deviation equal to $\displaystyle \frac{e^{2}}{h}$. Such effects are only observed when the phase is not destroyed by inelastic collisions, i.e. when the distance covered is lower than the inelastic mean free path. This is close to 0.1 $\mu$m at room temperature for GaAlAs-GaAs heterojunctions or in Si.}

\item{a one-dimensional quantum wire is analogous to a wave guide connected to electron reservoirs. When a voltage is applied between these reservoirs, electrons are injected in the wave guide. The number of one-dimensional channels for an electron depends on the number of energy levels below the Fermi energy. This number N of discrete levels due to the confinement perpendicular to the wire is fixed by the width of the quantum wire. Within these conditions the wire conductance is quantized and equal to $\displaystyle \frac{2Ne^{2}}{h}$.}

\item{discrete energy levels in different wells can interact through a tunnel effect. When a polarization is applied between the outer reservoirs, the conductance is small except when the quantum wells energy levels are aligned. Then the current is maximum and decreases for a further increase of the applied voltage. This gives rise to a negative differential resistance (see Fig. \ref{fig:fig2})}
\end{itemize}

These effects which limit the feasibility to reduce the device size below a certain value can also be used to invent new architecture. One electron devices are certainly the most surprising and promising effect. The simplest one is the tunneling junction shown on Fig. \ref{fig:fig3} \cite{ref6}. It is a metal-insulator-metal junction between two electron reservoirs. When the barrier width or height is large, the probability for an electron to cross the barrier when a polarization is applied is small and the tunneling is quantized: the electrons cross the barriers one by one. Then the system is equivalent to a capacitor C and an electron leak due to tunneling across the insulating layer (Fig. \ref{fig:fig3}.b). When a voltage $V$ is applied, the tunneling of an electron is possible if the energy difference

\begin{equation}
\Delta E =\frac{Q^{2}}{2C}-\frac{(Q-e)^{2}}{2C}
\label{eq:eqone}
\end{equation}

for the capacitor with a charge equal to $Q=CV$ and to $Q-e$ becomes positive. When, (i.e. $\displaystyle |V|<\frac{e}{2C}$) no electron can cross the barrier and we have a "Coulomb blockade". To maintain a constant current $I$, one must applied a saw tooth potential with a frequency $I/e$. It is difficult to realize such a system and the following electron box is much easier to make.
Let us take a metallic quantum box separated from two electron reservoirs on one side by an ideal capacitor $C_{s}$  and on the other one by a tunnel junction with a capacitance $C$  (Fig. \ref{fig:fig4}). Electrons can tunnel into the quantum box one by one until a charge $-Ne$, $N$  depending on the applied voltage. The simplest way to calculate $N(V)$  is to define the ionization level $\epsilon_{i}(N,N+1)$. For the box, $\epsilon_{i}(N,N+1)$  is equal to the total energy difference  $E_{TOT}(N+1)-E_{TOT}(N)$  which is roughly equal to  $\displaystyle \frac{dE_{TOT}}{dN}$. Applying the Koopmans theorem, this quantity is equal to the lowest one-electron energy level which can accommodate an extra electron when the box is already charged with $N$  electrons. A classical electrostatic calculation gives the potential of the charged box one must add to the HOMO energy level of the isolated box $\epsilon_{i,metal}$  to get the ionization potential:

\begin{equation}
\epsilon_{i}(N,N+1)=\epsilon_{i,metal}+\frac{e^{2}}{2(C+C_{s})}(N+\frac{1}{2}-C_{s}\frac{V}{e})
\label{eq:eqtwo}
\end{equation}

As shown on Fig. \ref{fig:fig4}.b, a stable -Ne box charge occurs when the metal Fermi level is located between $\epsilon_{i}(N,N+1)$  and $\epsilon_{i}(N-1,N+1)$. A step in $N(V)$  occurs each time the metal Fermi level is aligned with an ionization potential and we can easily control the box charge.
The next step is to realize a one-electron transistor \cite{ref6} shown on Fig. \ref{fig:fig5} with two tunneling junctions and an ideal capacitor. The gate potential  controls the charge of the dot and the current. More complicated devices allow to control the electrons one by one like the electron pump \cite{ref7} and the single electron turnstile \cite{ref8}. Experimental results and a good introduction can be found in Ref \cite{ref9,ref10,ref11}.
To observe a Coulomb blockade, the energy difference between two ionizations levels  $\displaystyle \frac{e^{2}}{2(C+C_{s})}$ as calculated from Eq. \ref{eq:eqtwo} must be larger than $kT$. The first experiments were done at low temperature but one must reduce the size to work at room temperature. Single electron electronics has been proposed by Tucker in 1992 \cite{ref12}. This would allow a further reduction of size and electrical power, both conditions being necessary to increase the speed of a device. Some components have already been realized \cite{ref13,ref14,ref15,ref16,ref17,ref18,ref19} but remain difficulties due notably to non-reproducibility.

\section{Molecular electronics}
If the reduction in size of electronic devices continues at its present exponential pace, the size of entire devices will approach that of molecules within few decades. However, major limitations will occur well before this happens, as discussed previously. For example, whereas in current devices electrons behave classically, at the scale of molecules, they behave as quantum mechanical objects. Also, due to the increasing cost of microelectronic factories, there is an important need for much less expensive manufacturing process. Thus, an important area of research in nanotechnology and nanoscience is molecular electronics, in which molecules with electronics functionality are designed, synthesized and then assembled into circuits through the processes of self-organization and self-alignment. This could lead to new electronics with a very high density of integration and with a lower cost than present technologies. In this section, we review recent progress in molecular electronics and we describe the main concepts at the origin of its development. We show that the latter is strongly connected to the invention of new tools to identify, to characterize, to manipulate and to design materials at the molecular scale. We also stress the importance of the basic knowledge which has still to be obtained in the way towards the integration of molecules into working architectures.

\subsection{Concepts and origins of molecular electronics}

Quite surprisingly, the first ideas of using specific molecules as electronic devices and of assembling molecules into circuits were proposed more than 30 years ago. At this time, the electronic processors were only in their infancy, and adequate tools to perform experiments on single molecules were not available. However, the concept of molecular electronics is appealing, and two components were proposed, the molecular diode and the molecular wire, which can be seen as elementary bricks to build more complex devices or circuits. In spite of these early proposals, practical realizations came only recently due to limitations in chemistry, physics and technology. In the following, we briefly describe the basic principles of the two components.

\subsection{Molecular diode}
In 1974, Aviram and Ratner \cite{ref20} proposed to make an electrical rectifier based on D-$\sigma$-A molecules between two metallic electrodes, where $D$ and $A$ are, respectively, an electron donor and an electron acceptor, and $\sigma$ is a covalent bridge. Figure \ref{fig:fig6} describes the physical mechanism for the rectification. The electronic states are supposed to be totally localized either on the D side or on the A side. The HOMO(D) and LUMO(D) are high in energy, compared to, respectively, the HOMO(A) and LUMO(A). Therefore, a current can be established at relatively small positive bias, such that the Fermi level at the A side is higher than LUMO(A), and the Fermi level at the D side is lower than HOMO(D), provided that the electrons can tunnel inelastically through the $\sigma$ bridge. Thus an asymmetric current-voltage $I(V)$ curve is expected like in a conventional electronic diode. In the prototype proposed by Aviram and Ratner, the donor group is made by a tetrathiofulvalene molecule (TTF), the acceptor group by a tetracyanoquinodimethane molecule (TCNQ), and the $\sigma$ bridge by three metylene bonds. These molecules are seen as the analogs of n and p-type semiconductors separated by a space-charge region. Twenty years have been necessary before the first experimental demonstration of rectifying effects in diodes based on molecular layers \cite{ref21}, and studies have been amplified in this direction recently \cite{ref22,ref23,ref24}. Nevertheless, the origin of the rectification is still matter of debate \cite{ref25}. In addition, the Aviram-Ratner principle at the level of a single molecule has not been demonstrated yet.

\subsection{Molecular wires}
The discovery of the first conductive polymers based on acetylene \cite{ref26,ref27} suggested that molecules could have interesting properties to make electrical wires at the molecular scale. These wires are necessary to connect molecular devices into circuits. In 1982, Carter \cite{ref28} suggested to address the acceptor and donor groups of an Aviram-Ratner diode with polyacetylene chains, and even to control several diodes in the same time. Using this technique, the expected density of components on a chip could be of the order of $10^{14}$ cm$^{-2}$, which is far beyond the possibilities of present technologies. One difficulty raised by Carter is that the transport in the molecular chains takes place in the form of solitons whose motion is slow. But this disadvantage would be largely compensated by the higher density.

\subsection{Molecular circuits}
Conventional microelectronic technologies presently follow a top-down approach, processing bulk semiconductor materials to make devices with smaller and smaller sizes. In molecular electronics, the basic components like molecular diodes and wires are used to build more complex devices and circuits. This is the bottom-up approach which starts from the molecular level to build a complete chip. For example, it has been recently proposed to create AND and XOR logical functions using assemblies of molecular films, molecular diodes and nanotubes (wires) \cite{ref29}. These functions could be used to design additioners and other operations which are currently done in CMOS technologies. However, the assembling of the elementary units will not necessarily lead to the expected device as the interaction between individual molecules may perturb their own function \cite{ref30}. Thus, other proposals suggest to consider the system as an ensemble and to use directly the chemistry to synthesize molecules with the required functions (e.g. additioners) \cite{ref31}. 

\subsection{Transport experiments on ensembles of molecules}
A great challenge of molecular electronics is to be able to transfer information at the molecular scale in a controlled manner. In the devices described above, it consists of a charged carrier (electron or hole) which is transmitted through a single molecule. During several decades, the impossibility to work at the level of a single molecule has been a major difficulty. However, the characterization of molecular ensembles has been undertaken by different means which we describe now.
\subsection{Molecules in a solvent}
The first approach consists to study the molecules in a solvent, and to probe the electronic transfer using a combination of chemical and physical methods. It is possible to characterize the electronic transport in solution through the measurement of the oxido-reduction by voltametry and of the electronic excitation by optical absorption. If the chemical reaction takes place inside a molecule, then reaction and intra-molecular transfer become equivalent.
Experiments have been made on donor-ligand-acceptor molecules, like mixed valence compounds recently synthesized \cite{ref32}. These organo-metallic complexes contain two metallic atoms with different oxidation degrees. The experiments allow the measurement of the electronic transfer between the two sites. For example, Taube has synthesized a stable molecule containing two ruthenium complexes which act as donor and acceptor groups \cite{ref33}. When the molecule is partially oxidized, a charge transfer is observed between the two ions. Other molecules have been made with a large distance of 24 \r{A} between the two ions, which is obtained by the intercalation of five phenyl groups \cite{ref34}. An electronic transfer is also observed in this system in spite of the long inter-site distance. This effect is attributed to the electronic coupling induced by the phenyl groups which play here the role of a molecular wire between the two ruthenium ions \cite{ref35}.
Interesting results have been also obtained on biological molecules. In the case of proteins \cite{ref36}, an increase of 20 \r{A} in the distance between the acceptor and donor sites leads to a decrease by a factor 10$^{12}$ of the transfer rate. Thus, in this process, proteins can be seen as a uniform barrier which limits the electronic tunneling. The importance of tunneling effects in biological molecules is presently a well studied topic.

\subsection{Molecular layers}
The second approach is the study of molecules which are self-organized in two-dimensional monolayers on the surface of a conductive substrate. Self-assembled monolayers (SAMs) are obtained by the Langmuir-Blodgett (LB) technique or by chemical grafting on a surface. The SAMs have to be well organized to avoid artifacts due to disorder. Electrical measurements require a second electrode which is made by evaporation of a metal on top of the SAM. This is usually the most difficult task because the metal must not diffuse into the SAM where it could make short circuits \cite{ref37}. In the same time, the contact resistance must be small which requires a good control of the metal/SAM interface.
The first results based on this technique have been published by Mann and Kuhn on LB films of alkane chains \cite{ref38}. Studies of the tunneling through the molecular films show that the molecules are good insulators. Similar results have been obtained on SAMs of n-alkyltrichlorosilane chemically grafted on a Si substrate \cite{ref39}. These chains realize a barrier of the order of 4.5 eV for the tunneling of electrons or holes \cite{ref40} and lead to a better insulating layer than SiO$_{2}$ at the nanometer scale.
Other works on SAMs concern molecular diodes in the sense of Aviram-Ratner: they are described in previous sections and in ref. \cite{ref21,ref22,ref23,ref24,ref25}. Also, nice results have been obtained by Fischer {\it et al} \cite{ref41} on organic heterostructures made with palladium phtalocyanines and compounds based on perylene. Using gold electrodes at 4.2 K, they obtain $I(V)$ curves with clear steps (Figure \ref{fig:fig7}) which are attributed to the resonant tunneling through the molecular levels. The $I(V)$ characteristics is symmetric when the heterostructure is symmetric, but it becomes rectifying when perylene layers are inserted on one side of the structure.

\subsection{ STM measurements on single molecules}
The invention of the Scanning Tunneling Microscope (STM) \cite{ref42,ref43} is a main step for the development of molecular electronics. The STM is based on a sharp tip (curvature radius in the nanometer range) which is placed at a small distance from a conductive surface in a way that electrons can tunnel from one electrode to the other. Using piezoelectric tubes, the position of the STM tip is fixed with an accuracy better than 1 \r{A}, horizontally and vertically. The STM is an efficient tool to study single molecules adsorbed on metallic or semiconductor surfaces \cite{ref44,ref45}. It can be used to image, to displace and to characterize single molecules.

\subsubsection{STM imaging of single molecules}
A natural application of the STM is the imaging of molecules adsorbed on a surface. In this mode, a constant bias is applied between the tip and the substrate. When the tip is displaced laterally, one measures the height of the tip which is adjusted in order to keep a constant tunneling current. The map of the height versus the lateral position is roughly representative of the topography of the surface, and thus gives information on the adsorbates. The adsorption of small molecules like CO and benzene on Rh(111) has been studied in detail \cite{ref46,ref47}, showing in some cases that self-organization can take place at the surface. Larger molecules like naphtalene or phtalocyanines have been also imaged \cite{ref48,ref49}. The STM allows to study the adsorption sites. Nevertheless, the interpretation of the image is not straightforward, as it does not give information on the atomic positions but on the electronic structure.
Most of the work on STM imaging of adsorbed molecules concerns metallic surfaces, and only few semiconductor surfaces. However, as microelectronics technology is based on silicon substrates, there is an increasing need to study organic molecules linked to a silicon surface \cite{ref50,ref51}. For example, Fig. \ref{fig:fig8} shows high resolution STM images of a Si(100) surface after deposition of thienylenevinylene tetramers, which belong to a new class of $\pi$-conjugated oligomers of particular interest as molecular wires \cite{ref52}. The silicon dimers, typical of a Si(100) (2$\times$1) surface, are visible, forming rows of grey bean shaped. On the top of these rows, bright features can be seen, with different shapes corresponding to different adsorption configurations \cite{ref51}. These results show that the molecules are quite conductive. Detailed studies allow a better understanding of the nature of the bonds between the molecules and the surface. Spectroscopic measurements are also possible, as detailed in the next sections.

\subsubsection{STM as a tool to manipulate and to fabricate molecular objects}
The STM tip is also a very interesting tool to manipulate the matter at the atomic scale. The forces (Van der Waals, electrostatic, chemical) between the atoms at the tip apex and the imaged object are used to displace atoms or molecules on a surface \cite{ref53}. The first controlled atomic manipulation was presented by Eigler {\it et al} \cite{ref54}. Xenon atoms were displaced on a nickel surface, by moving the tip which was kept close to the atoms. Artificial structures containing a small number of atoms were made using this technique. Xenon atoms were also transferred vertically from the surface to the tip in a reversible process \cite{ref55,ref56}.
The manipulation of molecules only came recently. Stipe {\it et al} have shown that it is possible to induce the rotation of acetylene molecules by excitation of C-H vibration modes with the STM \cite{ref57}. This work shows the importance of inelastic processes in the tunneling. This effect can be exploited to perform vibrational spectroscopy on individual molecules \cite{ref58} and is therefore a powerful technique to identify adsorbates, their chemical bonding and their local chemical environment. Due to the high spatial resolution of the STM, correlations between the electronic structure and vibrational excitation of adsorbed molecules can be determined. Moreover by using inelastic tunneling, molecules can be dissociated, desorbed or even synthesized, which is of essential importance in studies of molecular manipulation. Hla {\it  et al} have shown that the assembling of two molecules is possible using a STM \cite{ref59}. Starting from two C$_{6}$H$_{5}$I molecules, they first removed the iodine atoms. Then they approached the two phenyl groups and they observed the formation of a biphenyl molecule. This reaction, entirely realized using the STM, was made at 20 K whereas the conventional synthesis is impossible below 180 K. Recently, larger molecules have been manipulated, showing that their adsorption on a surface may induce important atomic reconstructions \cite{ref60}. The manipulation of specific parts of molecules may also lead to important changes in their $I(V)$ curve \cite{ref61,ref62}.
  
\subsubsection{STM spectroscopy of molecules}
In some conditions, the STM allows to study the electronic structure of molecules in connection with their interactions with the surface. In the spectroscopic mode of the STM, the tip is placed above an adsorbed molecule, and the current is measured as a function of the applied voltage. Thus, using this approach, a direct measurement of the transport properties of a single molecule is possible. Several studies have been applied to C60 molecules because they are quite stable and easy to manipulate due to their size. Joachim {\it et al} \cite{ref63} have measured the current through molecules on a gold substrate, at room temperature and with a small applied bias (50 mV). As the tip-molecule distance is reduced, the current increases at a very high rate which is interpreted by a distortion of the electronic levels due to the pressure induced by the tip. Other measurements at 4.2 K have been recently presented by Porath {\it et al} \cite{ref64,ref65}. C60 molecules are adsorbed on gold substrate covered by a thin amorphous carbon film. The $I(V)$ curves present clear steps which are interpreted by Coulomb blockade effects and resonant tunneling through the discrete states of the molecule. States close to the HOMO-LUMO gap are completely resolved in spectroscopy. The degeneracy of the HOMO and LUMO levels is broken, probably due to the tip-induced electric field or due to a Jahn-Teller effect.
Another way to probe the conductivity of a single molecule is to use slightly defective SAMs. A particular system has been mainly studied, with molecules terminated by a thiol end group. Self-assembly is routinely obtained on gold surfaces (and others) using chemical grafting based on sulfur-gold bonds \cite{ref66}. In a well-known experiment \cite{ref67}, dodecanethiol SAMs have been made with a small number of defects consisting of conjugated wires which are slightly longer than the dodecanethiol molecules. The dodecanethiol SAM forms a quite insulating layer \cite{ref38}. STM images of the surface show a smooth surface with only few bright spots at the position of the molecular wires. This result demonstrates that the molecular wires have a higher conductivity than the surrounding molecules. Other works have been performed in this direction \cite{ref68}. The spectroscopy of single molecules may be also evidenced by imaging the surface at different tip-sample bias. In the example presented in a preceding section where thienylenevinylene tetramers are adsorbed on Si(100), the images of the molecules are highly voltage dependent (Fig. \ref{fig:fig8}). At sufficiently high negative sample voltages, the molecules are visible, whereas at lower voltages, most of the molecules disappear. The interpretation is the following: at high voltages, electrons can tunnel through the HOMO of the molecule, whereas at lower voltages, only states close to the Fermi level of the semiconductor and associated with the Si dimers can contribute to the tunneling current. Thus, interesting developments with a STM are taking place in various areas. It remains that the experiments which have been done up to now are at the state-of-the-art level, and their transfer to create new technologies is not straightforward. If the STM is a very good tool to study the transport through a single molecule, it is clear that other approaches have to be developed to make molecular electronic devices.

\subsection{Molecules connected to nanoelectrodes}
Here we describe other approaches to determine the $I(V)$ characteristics of a single molecule using nanoelectrodes. Compared to STM, these methods lead to stable, permanent and symmetric junctions. Applications of these techniques are already foreseen in the field of chemical and biological sensors.

\subsubsection{Co-planar electrodes}
This approach is based on lithographic methods to make metallic electrodes separated by a small gap, of the order of 5-10 nm for the smallest ones. This system is used to characterize long molecules which are placed in such manner that they are connected to the two electrodes (Fig. \ref{fig:fig9}). The conductivity of carbon nanotubes has been studied using this approach \cite{ref69,ref70}. It has been verified that the nanotubes are metallic or semiconducting depending on their geometry. Carbon nanotubes are presently considered as the best candidates to make molecular wires, as the current can flow efficiently in long nanotubes ($\sim$ 1-0.1 mm). Co-planar electrodes are also used to measure the conductivity of DNA \cite{ref71}, and even to explore superconductivity effects in these systems \cite{ref72,ref73}.

\subsubsection{Nanopores}
In this approach, small holes are made in a thin silicon nitride film deposited on a metallic electrode. The pores are filled with the molecules. A second metallic electrode is evaporated on the top of the system to realize an electrical contact. Compared to techniques involving molecular films, a smaller number of molecules are probed in this approach ($\sim$ 1000). The number of junctions which are short circuited is also reduced \cite{ref74}. Measurements have been realized on 1-4 phenylene diisocyanide \cite{ref75} and [{\bf 1}: 2'-amino-4,4'-di(ethynylphenyl)-5'-nitro-1-benzenedithiol] \cite{ref76,ref77}. In the case of the first molecule, a symmetric $I(V)$ curve is obtained. In the case of the second one, it exhibits negative differential resistance with a high peak-to-valley ratio at low temperature (Fig. \ref{fig:fig10}). The origin of this effect could be related to the electronic structure of the molecule \cite{ref78}. In this case, the molecule could be used to realize oscillators.

\subsubsection{Breaking junctions}
In this third approach, a metallic wire (usually gold) is made by lithography on a flexible substrate. Then the wire is broken using a small mechanical perturbation. At the place where the wire is broken, there is a small gap of nanometer size where molecules can be deposited from a solution. Using piezoelectric tubes, the distance between the two electrodes is adjusted in such a way that some molecules make a bridge between the electrodes. One difficulty with this technique is that the number of molecules which are connected is unknown. Reed {\it et al} have studied the conductance of benzene-1,4-dithiol grafted on gold electrodes \cite{ref79}. Steps in the $I(V)$ curve are obtained, maybe due to Coulomb blockade effects. Other molecules have been studied \cite{ref80,ref81}, confirming for example that dodecanethiol is good insulator at small bias. Park {\it et al} have inserted a C60 molecule in a breaking junction \cite{ref82}. Using a gate voltage applied to the substrate, this system works as a transistor at the molecular level. In addition, a new effect is observed as the electronic transport is coupled to the motion of the C60 molecule. The results are explained by the oscillation of the molecule in the electrostatic field between two electrodes. Coupling to internal modes of vibration of the molecule is also suggested.

\subsection{Molecular circuits}
The fabrication of circuits based on molecules is obviously one main objective of molecular electronics. Even if much work is needed to develop the corresponding technologies, recent progress has been made in this direction. Collier et al \cite{ref83} have presented a molecular circuit based on the Teramac architecture \cite{ref84}. This computer architecture is defect-tolerant, meaning that it continues to work up to three percent of defected components. This kind of approach is ideal for molecular electronics since it will necessarily contain defective components and connections. Logic gates are fabricated from an array of configurable molecular switches, each consisting of a monolayer of electrochemically active rotaxane molecules sandwiched between metal electrodes. AND and OR functions have been realized. The circuit is quite equivalent to Programmable Read Only Memory (PROM). Its fabrication would be much cheaper than in CMOS technologies.
This work illustrates an important orientation of molecular electronics. It could be used to make logic devices at a relatively low cost. It is foreseen that expensive fabrication procedures in the electronic industry could be progressively replaced by techniques derived from molecular electronics to reduce fabrication costs. Thus, the main aim is no longer to realize nanometer scale devices to follow the first Moore's law, but to reduce costs which are presently continuously increasing (second Moore's law).

\newpage

\noindent{\bf CAPTIONS}\\

\begin{figure}[b]
\includegraphics[scale=.3]{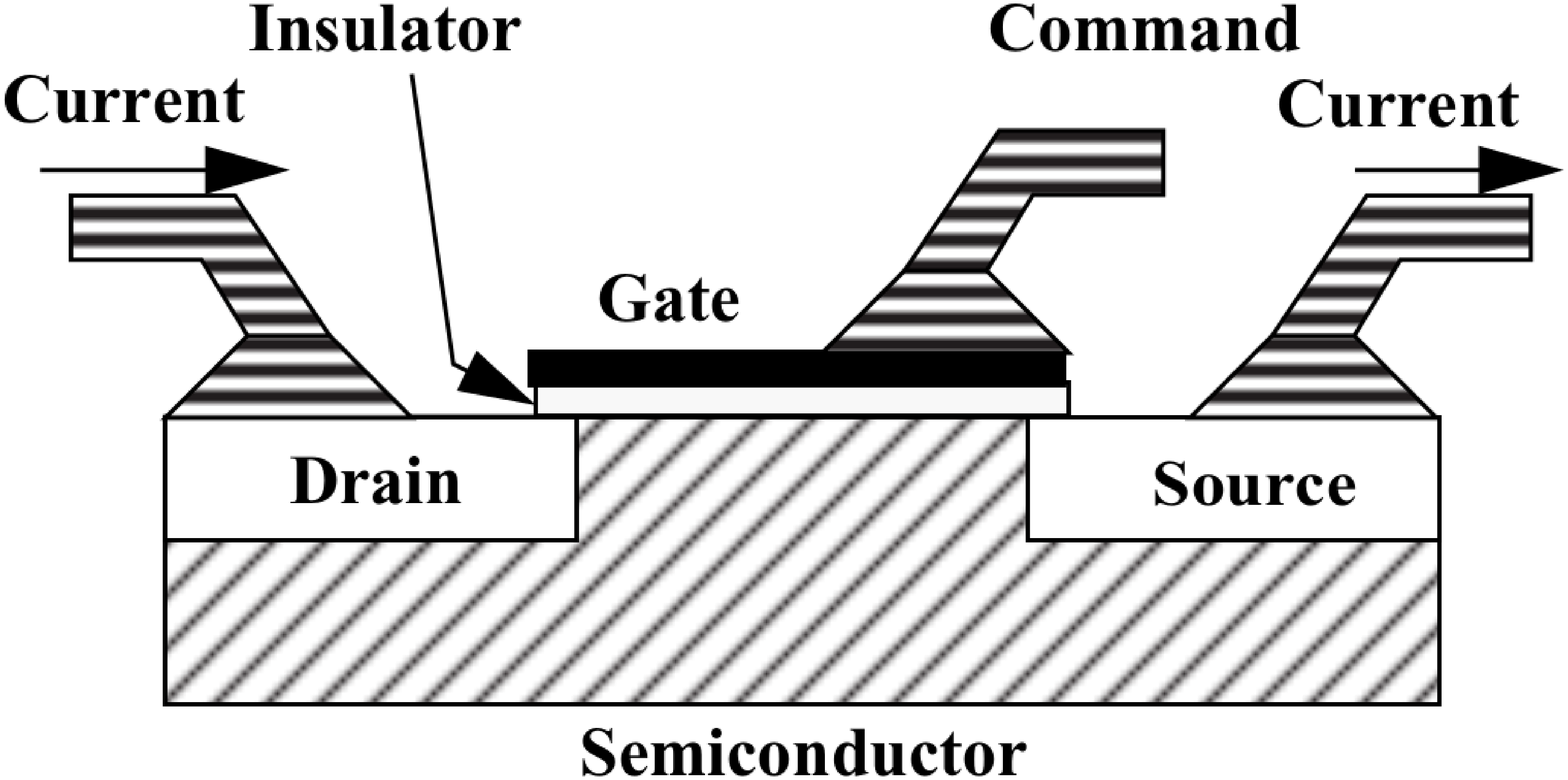}
\caption{\label{fig:fig1}}
\end{figure}

Figure \ref{fig:fig1}: A metal oxide semiconductor field-effect transistor (MOSFET). When the voltage of the gate is positive, electrons accumulate near the semiconductor surface making the channel between source and drain conducting.\\

\begin{figure}
\includegraphics[scale=0.75]{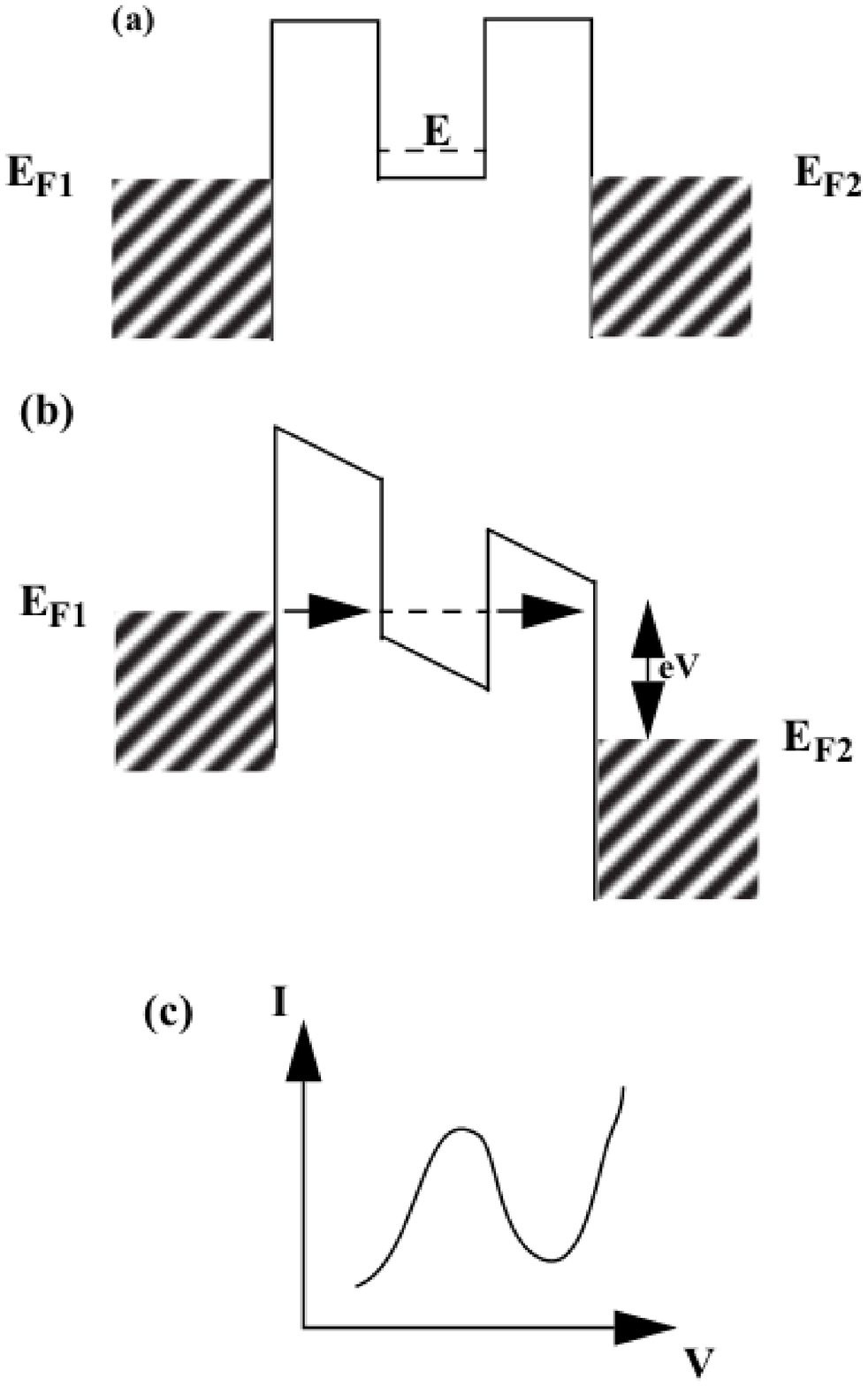}
\caption{\label{fig:fig2}}
\end{figure}

Figure \ref{fig:fig2}: Resonant tunneling effect (a) without and (b) with an applied voltage. (c) Current as a function of the voltage.\\

\begin{figure}
\includegraphics[scale=0.75]{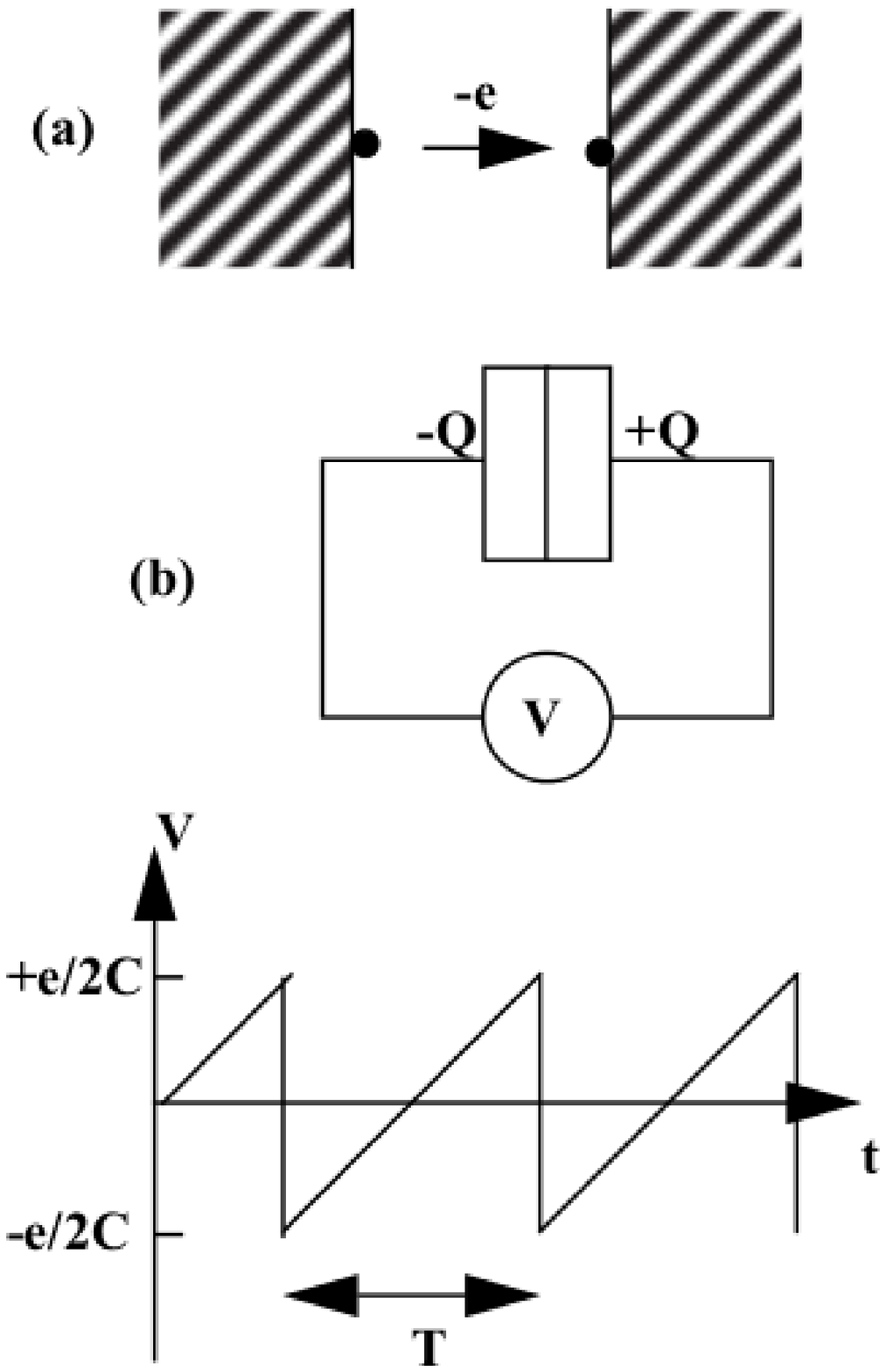}
\caption{\label{fig:fig3}}
\end{figure}

Figure  \ref{fig:fig3}: a) Metal - Insulator - Metal tunneling junction; b) Its equivalent diagram and c) the potential for a constant current I (=e/T) as a function of time.\\

\begin{figure}
\includegraphics[scale=0.75]{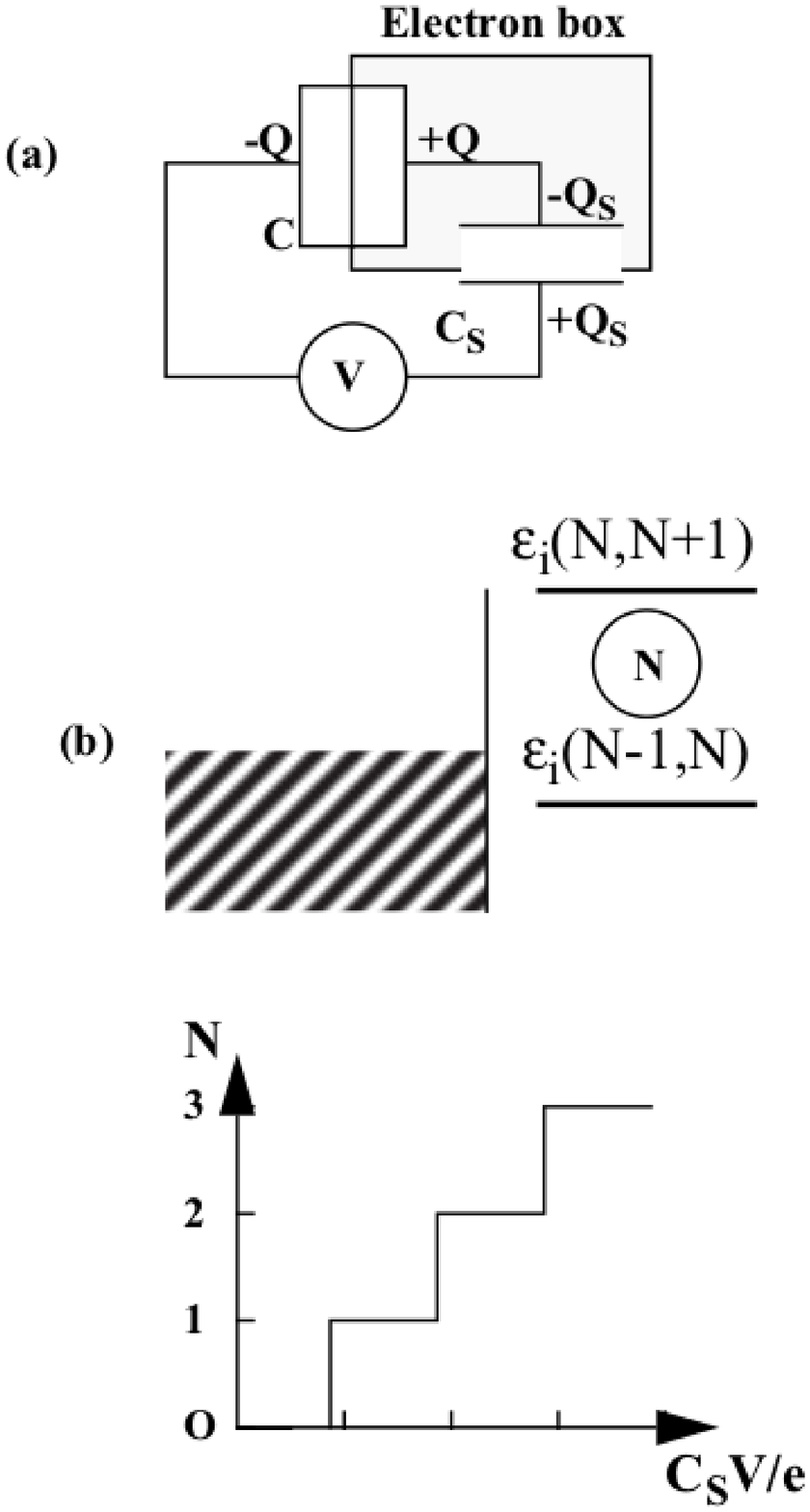}
\caption{\label{fig:fig4}}
\end{figure}

Figure \ref{fig:fig4}: a) Electron box (a) and its energy levels (b). $C_{s}$ is a tunneling junction. The electron number $N$ stored on the island varies as a staircase as a function of the applied voltage (c).\\

\begin{figure}
\includegraphics[scale=1.0]{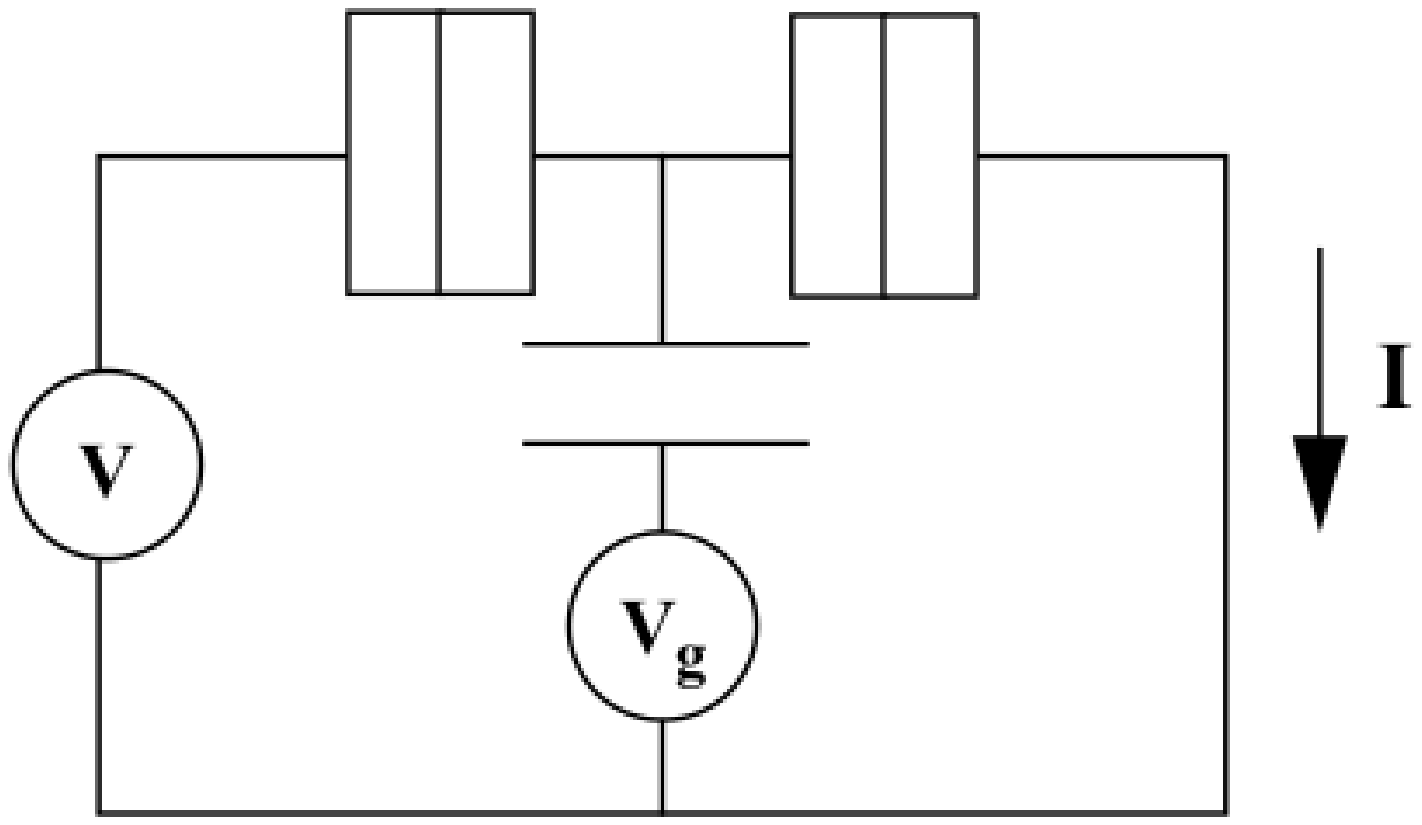}
\caption{\label{fig:fig5}}
\end{figure}

Figure \ref{fig:fig5}: One-electron transistor. The current $I$ and the charge of the island is controlled by the gate potential.\\

\begin{figure}
\includegraphics[scale=0.6]{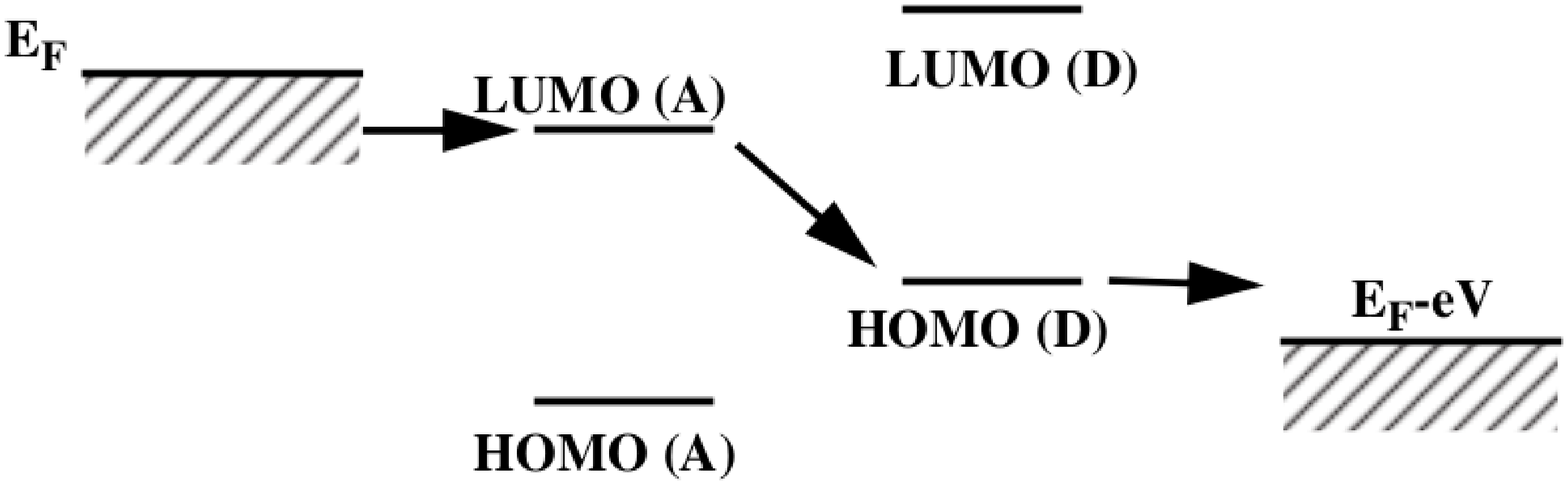}
\caption{\label{fig:fig6}}
\end{figure}

Figure \ref{fig:fig6}: The Aviram-Ratner mechanism for molecular rectification.\\

\begin{figure}
\includegraphics[scale=0.15]{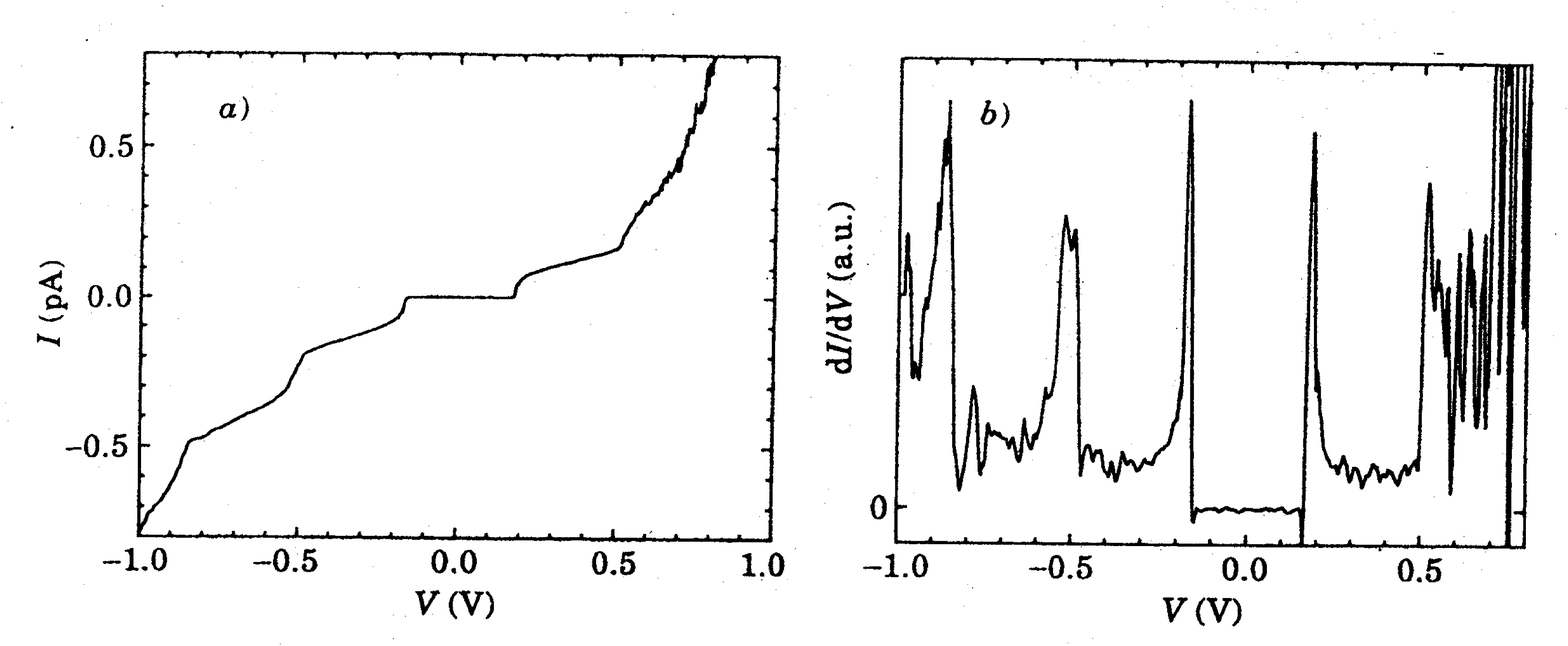}
\caption{\label{fig:fig7}}
\end{figure}

Figure \ref{fig:fig7}: $I(V)$ characteristics of a symmetric Au/Polymer/PcPd/Polymer/Au heterojunction measured at 4.2 K (a) and the corresponding derivative (b)); PcPd is an octasubstituted metallophtalocyanine (from ref. \cite{ref41}).\\

\begin{figure}
\includegraphics[scale=1.0]{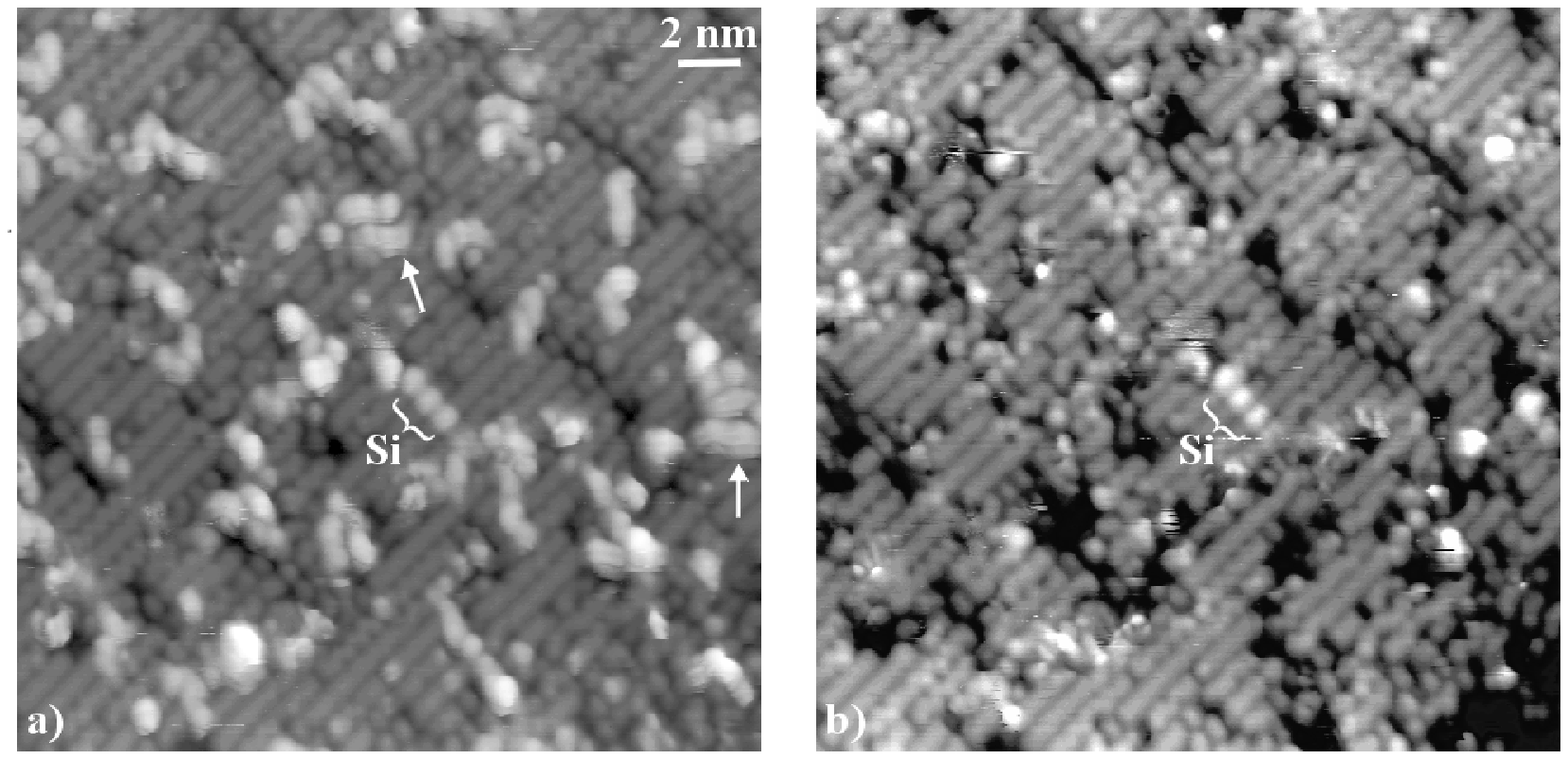}
\caption{\label{fig:fig8}}
\end{figure}

Figure \ref{fig:fig8}: Voltage dependent STM images of the Si(100) surface after deposition of 4TVH oligomers. The sample bias was in (a) -2.1 V and in (b) -1.3 V (from ref. \cite{ref51}).\\

\begin{figure}
\includegraphics[scale=0.5]{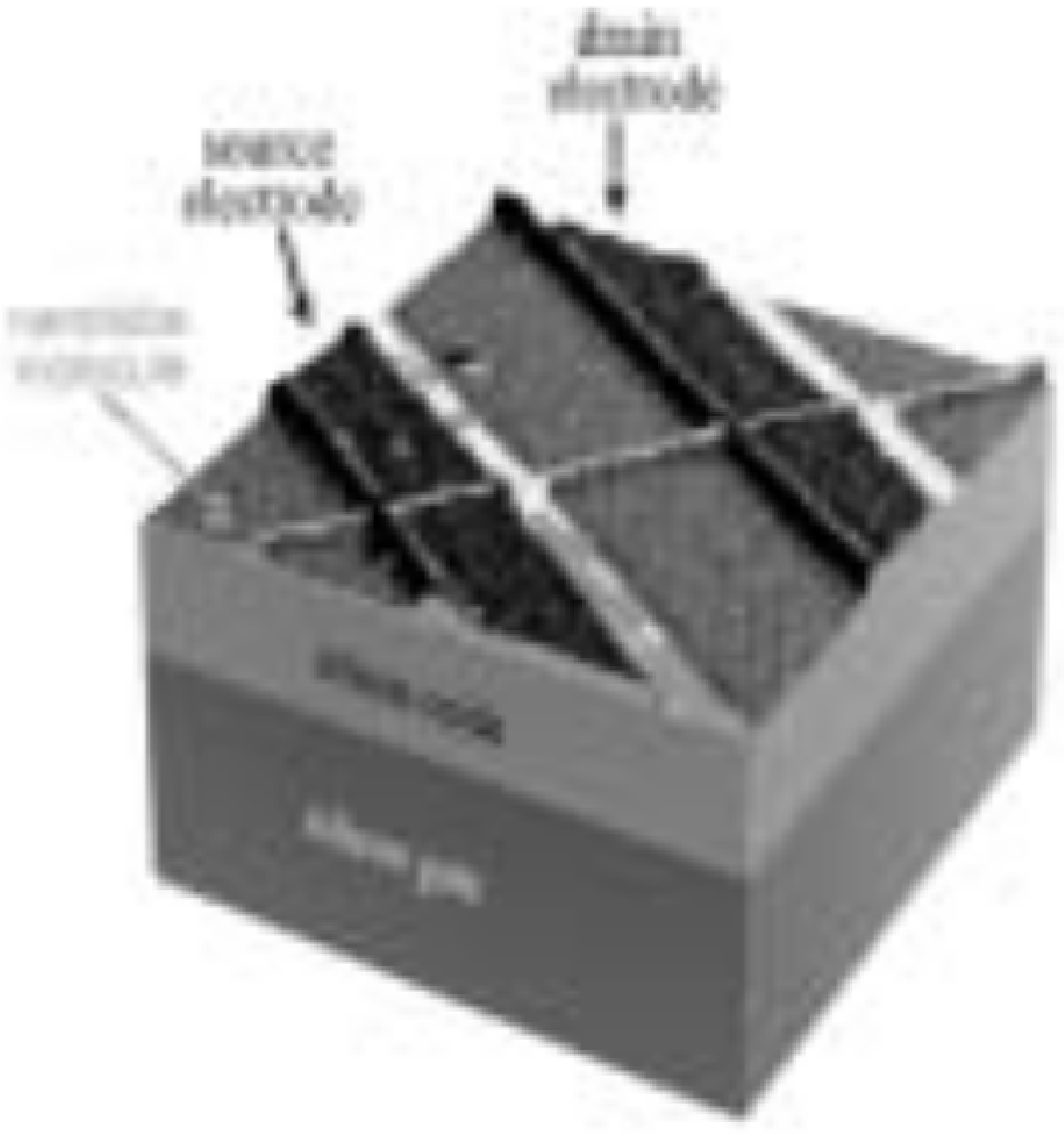}
\includegraphics[scale=0.33]{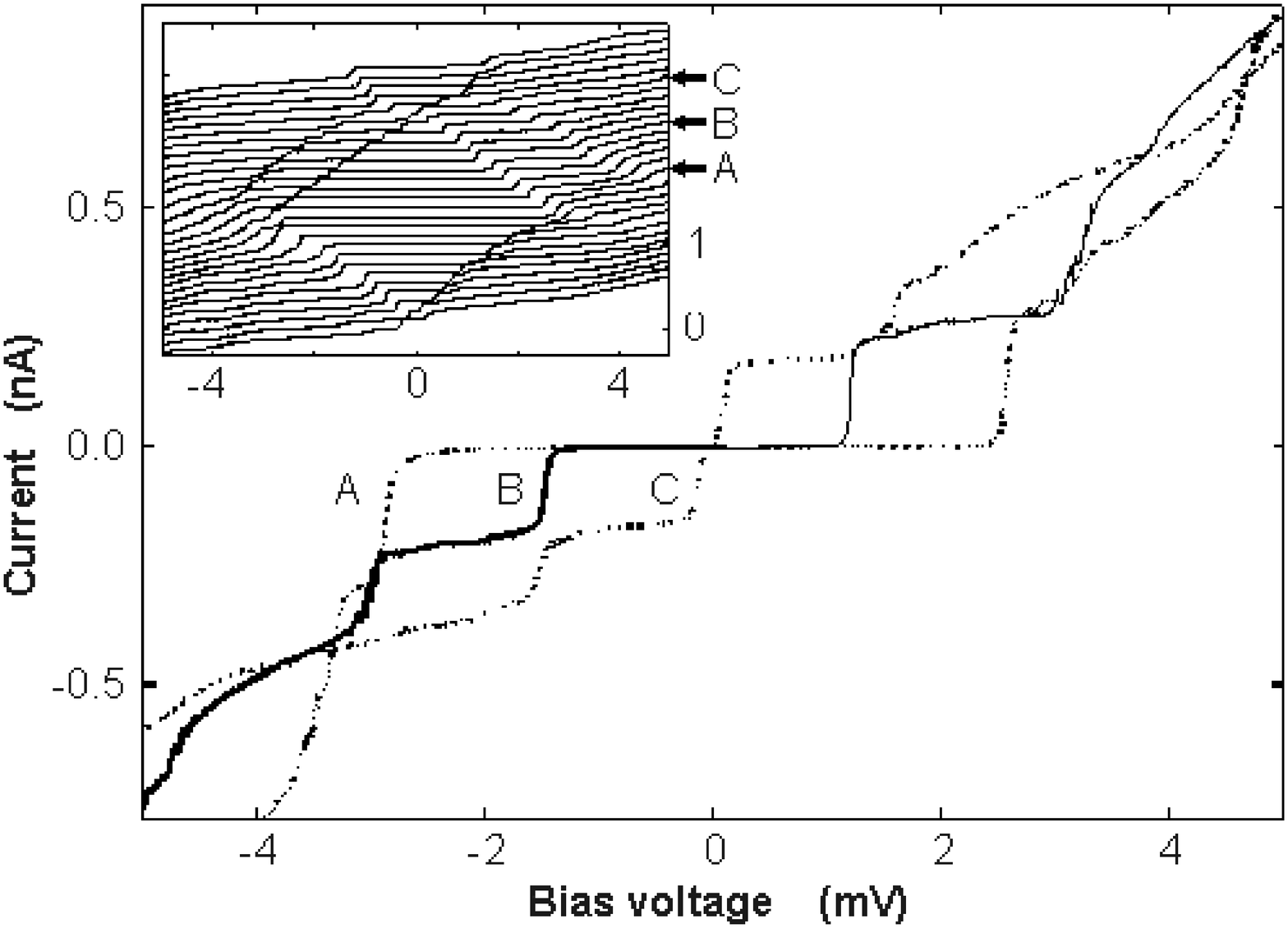}
\caption{\label{fig:fig9}}
\end{figure}

Figure \ref{fig:fig9}: Nanotubes on co-planar electrodes and corresponding $I(V)$ characteristics at different gate voltages (courtesy: C. Dekker).\\

\begin{figure}
\includegraphics[scale=0.5]{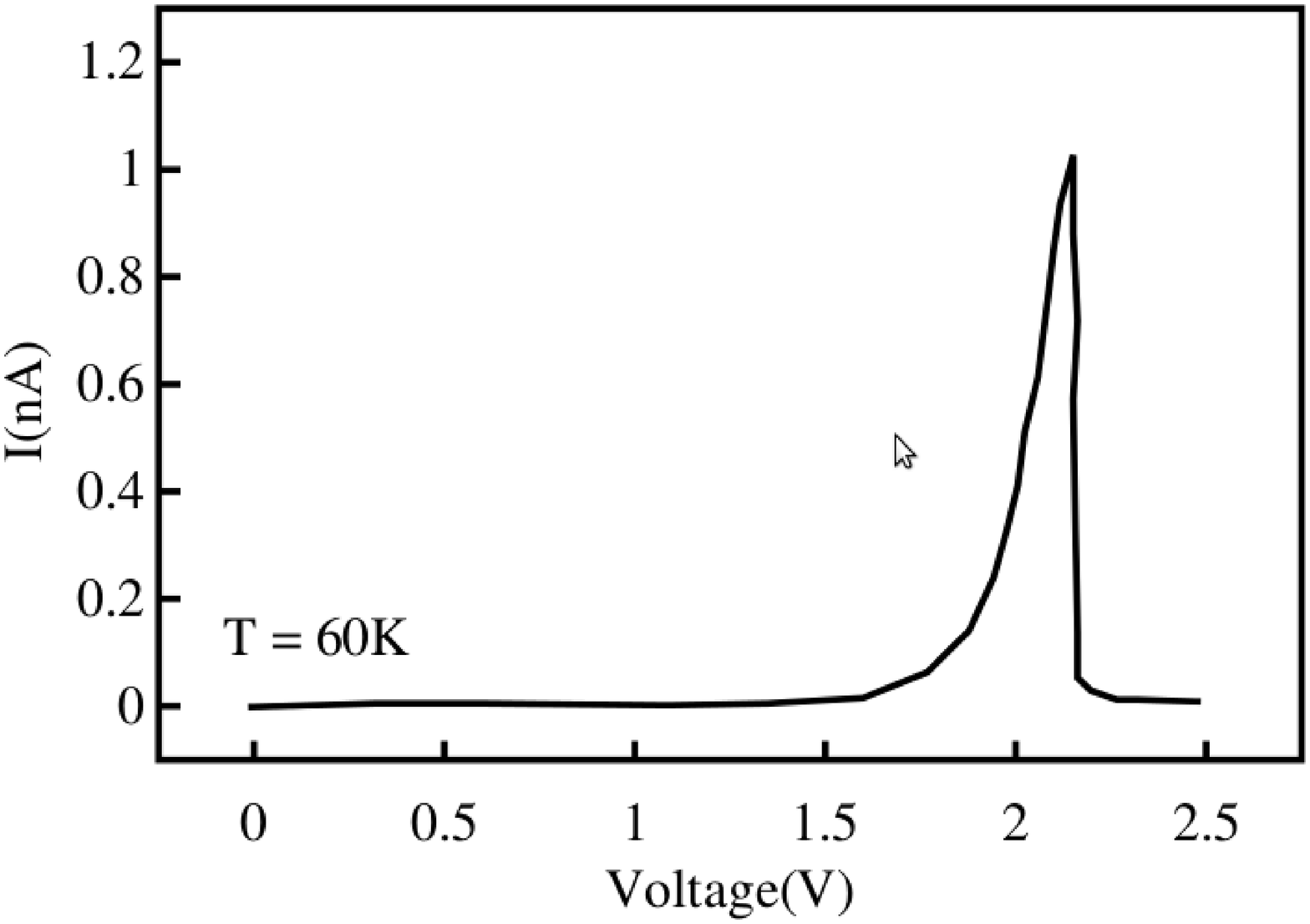}
\caption{\label{fig:fig10}}
\end{figure}

Figure \ref{fig:fig10}: $I(V)$ characteristics of a Au-({\bf 1})-Au device at 60 K (from ref. \cite{ref77}).\\

\noindent{\bf FIGURES}\\

\vspace{10cm}

\vspace{7cm}

\vspace{10cm}

\vspace{10cm}

\end{document}